# Asymmetric double-pulse interferometric frequency-resolved optical gating for visible-wavelength time-domain spectroscopy

Yi Chan[1], Fu Deng[1] and Jingdi Zhang[1*]

[1]Department of Physics, The Hong Kong University of Science and Technology, Kowloon, Hong Kong SAR, People's Republic of China
*Correspondence: jdzhang@ust.hk

**Abstract:** Ultrafast science and technology have brought in burgeoning opportunities to optical metrology, strong-field physics, non-equilibrium physics, etc., through light-matter interaction due to ever-advancing temporal resolution and peak power of ultrafast laser. The superior temporal and spectral resolution, has brought forth pump-probe spectroscopy for ultrafast dynamic study of transient states in various intriguing materials, such as quantum materials, metamaterials, and plasmonic materials, by directly reporting spectroscopic complex response function, i.e., $\tilde{\epsilon}(\omega)$ using either time- or frequency-domain-based probes. In contrast to its frequency-domain counterparts, e.g., FTIR and ellipsometry, *time-domain spectroscopy* outstands by providing not only superb spectroscopic phase sensitivity but also exceptional temporal resolution due to its pulsed nature. To extend detection range of *time-domain spectroscopy* into the challenging visible frequencies, we propose an *interferometry-type* frequency-resolved optical gating (FROG). Our numerical simulation shows, when operating in a double-pulse scheme, a unique phase-locking mechanism can be activated, and therefore preserves both zero- and first-order phases ($\varphi_0, \varphi_1$), that are otherwise inaccessible to standard FROG measurement. Followed by time-domain signal reconstruction and analysis protocol, we show that *time-domain spectroscopy* with sub-cycle temporal resolution is enabled and well suits the need of ultrafast-compatible and ambiguity-free method for complex dielectric function measurement at visible wavelengths.

Time-domain spectroscopy (TDS), referring to a complete waveform mapping of oscillating electromagnetic pulse with sub-cycle temporal resolution, is an increasingly important tool for measurement of complex optical response functions, i.e., dielectric function $\tilde{\epsilon}(\omega)$, optical conductivity $\tilde{\sigma}(\omega)$ or refractive index $\tilde{n}(\omega)$, particularly at far-IR and mid-IR wavelengths. Ultrafast laser, emitting pulsed electromagnetic wave of femtosecond and even attosecond duration, renders remarkable temporal resolution and detection bandwidth, and, therefore, has been widely used for dynamic study of non-equilibrium states in complex materials [1–4], referred to as *pump-probe spectroscopy* [5]. Through light-matter interaction, an intensive pulse, known as *pump* beam, initiates non-equilibrium dynamics and is followed by a weaker pulse, known as *probe* beam, that reports on subsequent transient state at a delayed time [6,7]. One preeminent probing method, known as time-domain spectroscopy, has been increasingly important for such endeavor, without relying on Kramers-Kronig analysis that is otherwise required by rival frequency-domain technique, such as broadband FTIR and standard reflectance or transmittance measurement. Time-domain spectroscopy typically operates at far- and mid-IR range, i.e. THz and multi-THz frequencies, and maps out waveform of a target pulse $E(t)$ with sub-cycle resolution by convolving its relatively slow oscillating field and another much shorter gate pulse via electro-optic sampling (EO-sampling). With proper referencing, this method completely registers all necessary spectroscopic amplitude and phase information encoded into the time-domain signal, indispensable for an ambiguity-free extraction of complex optical response functions within spectral range of the probe pulse [8]. Although standard TDS is a powerful spectroscopic technique at far- and mid-infrared wavelengths, the lack of easy access to sub-femtosecond pulse, compatible with standard EO-sampling crystal, sets cut-off detectable wavelengths to near-IR ($\lambda$<3 μm), prohibiting effective TDS measurement in the visible range [9]. As such, to date, most research efforts of ultrafast optical study at near-infrared, visible and ultraviolet wavelengths have been restricted to delivering dynamic spectra only in magnitude by reflectance, transmittance, or absorbance measurements, but with limited information on phase. In particular, ultrafast spectroscopic attribute of advanced materials, e.g., electronic transition and its coupling to other degrees of freedom in this regime, can be only inferred but not fully resolved. Therefore, it is beneficial to design and implement experimental tools, capable to provide both intensity and phase sensitivity at shorter wavelengths–near-IR, visible and even UV–on ultrafast timescales toward complete understanding of novel photo-induced transient states and the pertinent dynamics. To obtain phase information in optical regime, spectral interferometry (SI) was first introduced to extract the spectroscopic phase change of transient states of solid-state materials by a pair of identical pulses (reference and probe) with a delay such that pump pulse is temporally encaged in between [10–12]. Although temporally separated, the coherent pulse pairs are spectrally overlapped and interfered to generate wavelength-dependent spectral fringes, of which displacement in frequency domain is used to quantify pump-induced dynamic phase changes in the sample. Later on, TADPOLE was proposed and demonstrated to further increase the sensitivity of SI after being combining with second-harmonic generation frequency-resolved optical gating (SHG-FROG) [13]. It emerged as a prominent method for complete retrieval of both amplitude and phase of an unknown ultraweak pulse on femtosecond or even shorter timescale due to its self-referencing and nonlinear characteristics [13–15]. Although SI is simple in principle, it faces some practical issues owing to adverse effect of imperfect collinear alignment and spatial modes matching, as well as compromised spectral resolution, approximately five times worse than instrumental resolution of spectrometer being used [16]. Besides spectral interference, ultrafast Brewster-angle spectroscopy (BAS) at multiple angles can, as well, provide phase sensitivity at ultraviolet-visible wavelengths, supplemented by Fresnel's equation for inverse calculation of complex dielectric function [17,18], for instance, in transient state of GaAs [18,19]. Despite its success, the requirement on setting incident angle to the wavelength-dependent Brewster angle brings inherent difficulty and uncertainties in any broadband measurement, and the issue can be further compounded in presence of a strong resonance that drastically modifies Brewster angle upon changing wavelengths. Moreover, requirement on large and varying incident angles become unfavorable to cryogenic experiments involving vacuum chambers.

As such, it is desirable to develop new and inexpensive method capable to retrieve both spectroscopic amplitude and phase information, in particular, at visible wavelengths with high temporal and spectral resolution. To circumvent the abovementioned challenges, we propose to directly retrieve and completely reconstruct time-domain signal of femtosecond optical pulse with hyper temporal resolution (<0.1 fs) by resorting to the interferometry-type FROG spectroscopy. We show the method, being interferometric, activates a phase-locking mechanism, resulting in extended sensitivity to dispersion of optical pulse to the lowest order and, therefore, is free from phase ambiguity, commonly suffered by standard FROG retrieval.

In principle, FROG alongside auxiliary retrieval algorithm is the only available method for hyper-resolution measurement of amplitude and phase on sub-femtosecond timescale, regardless of complexity of optical pulse concerned [20,21]. Among various FROG-based technique, SHG-FROG is the most suitable for measuring and retrieving weak pulses due to its simplicity and outstanding susceptibility. However, conventional single-pulse FROG retrieval is inhabited by three inherent ambiguities in a reconstructed pulse: time direction, zero-order absolute phase ($\varphi_0$) and first-order translational phase ($\varphi_1(t)$). These ambiguities root in the fact that FROG is essentially an intensity spectrogram in the time-frequency domain, only registering the chirp of laser pulse, i.e., second- or higher-order phase ($\varphi_n(t), n \geq 2$). Removing these ambiguities has no immediate gain for standard pulse characterization, but is indispensable for extracting anything of spectroscopic significance, e.g., complex optical response function.

To restore the missing time direction and low-order phases, we propose a design of interferometric SHG-FROG spectroscopy, operating in an asymmetric double-pulse scheme (Fig.1). The double-pulse train is comprised of a *probe* pulse to interact with specimen by reflection and a *calibration* pulse to activate the phase-locking mechanism yet to be detailed below. By asymmetric, we mean that, despite spectrally overlapping, the constituent pulses are neither the same in duration nor in peak intensity, by setting calibration pulse to be longer and weaker than probe pulse. For optimal result, *calibration* pulse is set to lead the *probe* pulse at fixed time interval, $T$, and possess a longer duration, i.e., linear chirped, after traversing a stretcher. The total field of double-pulse train, $E(t)$, follows the superposition of calibration pulse, $E_{cal}(t) = a_{cal}(t)e^{i(\omega_0 t + \varphi_{cal})}$, and delayed probe pulse $E_{pr}(t) = a_{pr}(t)e^{i[\omega_0(t-T)+\varphi_{pr}]}$ where $a_{cal}$ ($a_{pr}$) is calibration (probe) pulse envelope, $\omega_0$ carrier wave frequency and $\varphi_{cal}(\varphi_{pr})$ time-dependent phase expandable to all orders. Double-pulse train could experimentally be synthesized by combing the two coherent fields at beam splitter into a unity, and then dispatched to a standard SHG-FROG, that splits total field $E(t)$ into two replicas of $E'(t)$ and $E'(t-\tau)$ at variable delay $\tau$ and generates FROG signal filed ($E_s(t,\tau)$) inside nonlinear crystal ($\beta$-BBO). The signal is finally spectral analyzed for a 2D spectrogram $S_{FROG}(\omega,\tau)$, referred to as FROG trace that follows [13]

$$S_{FROG}(\omega,\tau) \propto \left| \int_{-\infty}^{+\infty} E'(t)E'(t-\tau)e^{-i\omega t}dt \right|^2 \qquad (1)$$

Unlike conventional single-pulse FROG trace, the double-pulse field generates a quasi-triple-peak FROG trace with spectral interference fringes (Fig. 1c), resembling the multiple-pulse interferometric FROG (MI-FROG) [22]. However, in MI-FROG trace, the interferometric attribute is pronounced as horizontal-fringes modulated central peak only, whereas our temporal asymmetric double-pulse train notably generates diagonally sliding fringes, interconnecting side and main peaks in FROG trace, indispensable for activating phase-locking mechanism that preserve phase information of optical pulse at all orders.

Insights can be obtained by decomposition analysis without and with effect of spectral interference. Without spectral interference, the quasi-triple-peak FROG trace of asymmetric double-pulse train is comprised of three major parts shown in Fig. 2, (see Supplementary for details). The central peak results from intensity sum of FROG signals of self-interacting calibration and probe pulse (Fig. 2 b, c) near zero delay ($\tau=0$, add inset) in the absence of cross interaction of one another. Because calibration and probe pulse enter FROG in pair at interval of T, they spectrally interfere to generate horizontal fringes, modulating central peak of the FROG trace (Fig. 2e), as in MI-FROG. Note that the left (right) side peak (Fig. 2d) originates from cross interaction between calibration and probe pulse $E_{cal}(t)$ (or $E_{pr}(t-T)$) and $E'_{pr}(t-\tau-T)$ (or $E'_{cal}(t-\tau)$) by scanning delay $\tau$ proximal to $-T$ ($+T$). Although symmetric central peak alone still obscures time direction, the associated ambiguity is readily removed by referring to directional side peaks, i.e., Λ-shaped (or V-shaped) pair of side peaks, resulting from gating effect of short *probe* pulse on linear-chirped long *calibration* pulse.

We highlight that, in Fig. 2f, besides horizontal fringes overlapping central peak, a new pair of diagonal fringes come to light when delay between replicas of double-pulse train $\tau$ permits *calibration* pulse in one train, $E_{cal}(t)$, to simultaneously overlap with both *calibration* and *probe* pulses in the other, $E'_{cal}(t-\tau) + E'_{pr}(t-\tau-T)$ (Region B, in Supplementary Fig. S1), i.e., a double-pulse train temporally bridged by a third pulse. This bridge configuration generates phase-locked double-pulse FROG signals in accordance to Eq. 1, containing three terms: self-interaction calibration pulse $S_{cal}(\omega,\tau)$, cross-interaction $S_{cal-pr}(\omega,\tau,T)$ and interference $2\sqrt{S_{cal}(\omega,\tau)}\sqrt{S_{cal-pr}(\omega,\tau,T)}\cos(\Delta\varphi_0 - \varphi_f)$. $T$ is expressed out explicitly to denote calibration-probe pulse interval, $\Delta\varphi_0 = \varphi_{0_{cal}} - \varphi_{0_{pr}}$ denotes absolute phase (zero order) difference between calibration and probe pulses, and $\varphi_f$ denotes algorithm-retrievable higher order phase difference between complex electric field corresponding to $S_{cal}$ and $S_{cal-pr}$ (see Supplementary S2). In stark contrast to interference term in conventional SI measurement, in which phase in the sinusoidal interference term is determined by two coupled variables $T$ and $\Delta\varphi_0$, phase in interference term in our case is single-variable dependent. The former makes it inextricably complicated to converge toward a unique solution, as it is a 2D search, whereas the latter is reduced to 1D problem with a unique solution in practice. It is this term predominantly activates the phase-locking mechanism. We stress that, as now $T$ is decoupled from sinusoidal modulation term and absorbed into magnitude part of the FROG spectrogram, task of determining first-order dispersion ($\varphi_1$) is now naturally taken on by FROG retrieval algorithm, due to the calibration functionality offered by calibration pulse field $E_{cal}(t)$ in the replica train, rather than referring to interference fringes as in other interferometric technique. Of primary importance is that the *phase-locking mechanism* only locks the difference of zero-order phases in *calibration* and *probe* pulse and precluded directly yielding anything quantitative. Precise determination of phases is left to the mighty FROG retrieval algorithm with a guaranteed minimal uncertainty. For this reason, in the new scheme, individual absolute phase ($\varphi_{0_{cal}}$ & $\varphi_{0_{pr}}$) still remains ambiguous. Nevertheless, their difference $\Delta\varphi_0$ is now completely ambiguity-free. In short, although named *interferometric FROG*, it is fundamentally distinct from standard spectral interferometry, in the sense that interferometry herein only establishes the phase-locking mechanism but are precluded contributing to report on actual value of phase, on which the work is fulfilled by non-interferometric but extremely accurate FROG part of the instrument. In turn, by properly allocating the workload, we leverage advantages and avoid disadvantages of interferometry and FROG measurement for the best outcome. This nontrivial extension of sensitivity to lowest order phase allows for full reconstruction of time-domain signal of pulse field that is essential for accurate extraction of both spectroscopic amplitude and phase information from a specimen.

To demonstrate full spectroscopic capability of asymmetric double-pulse interferometric FROG (ADI-FROG) in the visible regime, on the numerical simulation level, we examine two semiconductor materials $WS_2$ and modulated-GaAs (bandgap red-shifted by 285 nm), in both static and non-equilibrium states. In all cases, short probe pulse is first interacted with sample in reflection, dictated by Fresnel equation, and combined with linearly-chirped calibration pulse of longer duration and slightly lower peak intensity, so as to synthesize the asymmetric double-pulse train in preparation for succeeding phase-sensitive interferometric FROG measurement. In practice, we suggest adopting a twin noncollinear optical parametric amplifier (twin-NOPA) for generating CEP-interlocked probe and calibration pulse. Finally, the double-pulse train is sent to SHG-FROG for generating FROG signal beam, to be spectrally resolved at different delay $\tau$ for 2D FROG trace $S_{FROG}(\omega, \tau)$. The same process is repeated for taking a reference scan without interacting probe pulse with sample, but with metal surface, e.g., silver, with unity reflection.

FROG traces of reference and sample scans are computed using Eq. (1) (Fig. 3a, b), treated as experimental results in this work, and subsequently fed to FROG retrieval algorithm based on generalized projections, established by DeLong, Taft and Trebino [23,24]. At the simulation level, a retrieval with corresponding G-error of $10^{-4}$ can be routinely achieved in reference to the input "experimental" trace, signaling satisfying level of convergence toward a high-quality retrieval [25]. We simultaneously obtain retrieved time-dependent amplitude and phase of calibration-probe double-pulse trains in two cases: reflecting probe off a target sample (red) and a high-reflective reference (blue) (Fig. 3c, d). By reference, we mean a probe pulse reflected by surface with unity reflectance, e.g. gold or silver, and by sample, a probe pulse reflected by either $WS_2$ or modulated-GaAs in this work. As the FROG retrieval algorithm outputs one *intensity curve*, that is only normalized and comes with random offset in absolute time, and another *phase curve*, that sits on random absolute-phase baseline, these retrieved pulse trains have to be manually displaced in time and re-normalized in amplitude, benchmarked by universally identical calibration pulses. Specifically, to capture *group delay* in probe pulse caused by first-order dispersion $\varphi_1(\omega)$ in the sample, retrieved time-dependent intensity and phase profiles of sample scans (red) are "realigned" to reference scans (blue) until perfect overlap between respective leading calibration pulses is achieved. Although the exact value of the absolute phase $\varphi_0$ still remains unknown, the abovementioned phase-locking mechanism guarantees a fixed zero-order relative phase between calibration and probe pulses within each double-pulse train, i.e., reference $\Delta\varphi_{0_{cal-ref}}$ or sample $\Delta\varphi_{0_{cal'-sam}}$. In turn, any phase difference between the two phase curves, that survived the calibration-pulse benchmarking process, can faithfully reflect the quested spectroscopic phase factor introduced by the sample to a pristine probe pulse, enabling a complete reconstruction of time-domain signals of both sample (red) and reference (blue) pulse (Fig. 3e). Reconstruction of time-domain signal is done by exclusively isolating out time-domain amplitude $a(t)$ and phase $\varphi(t)$ of trailing probe pulses (shaded regions in Fig. 3c and 3d), subsequent zero padding remaining empty space (Fig. 3e) and applying equation $E_{pr}(t) = a(t)e^{i(\omega_0 t + \varphi(t))}$, where $\omega_0$ is carrier wave frequency, a known parameter. Finally, as standard treatment in time-domain spectroscopy, we Fourier transform reconstructed sample and reference scans into frequency domain for $\tilde{E}_{sam}(\omega)$ and $\tilde{E}_{ref}(\omega)$ and then directly calculate for spectroscopic complex dielectric function $\tilde{\epsilon}(\omega)$ in frequency, or equivalently $\tilde{\epsilon}(\lambda)$ in wavelength, using inverse Fresnel equation:

$$\tilde{\epsilon}(\omega) = \frac{1}{2} + \frac{1}{2}\left[\frac{1-\tilde{E}_{sam}(\omega)/\tilde{E}_{ref}(\omega)}{1+\tilde{E}_{sam}(\omega)/\tilde{E}_{ref}(\omega)}\right]^2 \qquad (2)$$

$\tilde{\epsilon}(\omega) = \epsilon_1 + i\epsilon_2$, where $\epsilon_1$ and $\epsilon_2$ are real and imaginary part of the complex-valued dielectric constant, respectively.

To verify feasibility of the method described above, we performed example numerical simulation on retrieval of spectroscopic complex dielectric function of two canonical semiconductors (WS$_2$ and GaAs) in the visible range (550-750 nm). Figure 4a shows the retrieved equilibrium state dielectric function of WS$_2$–$\epsilon_1$ (λ) and $\epsilon_2$ (λ) (solid lines)–using the interferometric FROG augmented time-domain spectroscopy, in comparison to that given by analytical model (dashed lines). The capability of interferometric FROG spectroscopy is further witnessed by success on GaAs, a system more difficult to measure due to its significantly broader inter-band transition peak, i.e., rapidly damping feature in the reconstructed time-domain signal. To raise the methodology close to its full potential, we purposely redshifted the transition band near X-point of GaAs by 260 nm into spectral range of the probe pulse (500 - 800 nm). Amid the artificially introduced spectroscopic complexity, complete dielectric constant can still be successfully retrieved in case of both WS$_2$ and GaAs, independent of details of the transition resonances and the pertinent linewidth. Moving toward the ultimate goal of non-equilibrium dynamic study in the visible range, we now verify the effectiveness of the method in investigation of photo-induced transient state of either WS$_2$ or modulated-GaAs after manually adding Lorentz oscillator or shifting resonance frequency. As expected, the resultant sample probe pulse shows an even more complicated waveform as a result of the emergence of additional oscillator in the visible range. Nevertheless, both $\epsilon_1$ (blue) and $\epsilon_2$ (red) of WS$_2$ (Fig. 4c) and GaAs (Fig. 4d) in respective non-equilibrium states are accurately retrieved, showing all spectroscopic resonance features, in perfect agreement with analytical model (dashed line).

In conclusion, the proposed method demonstrates the interferometric FROG measurement can facilitate time-domain spectroscopic study owing to the extended sensitivity to low-order phases of femtosecond laser pulse, as evident by numerical simulation results herein. Such methodology can shift the spectral barrier for standard time-domain spectroscopy beyond visible wavelengths, and can be applied to other wavelengths as well, e.g., near-IR. By fully capturing amplitude and phase information on ultrafast timescale, in general, it would open up new opportunities for elucidation of novel photo-induced phenomena and demonstration of all-optical manipulation in advanced materials.

**Acknowledgement:** The authors would like to acknowledge support from Hong Kong Research Grants Council (Project NO. ECS26302219) and HKUST Research Equipment Competition (REC19SCR14).

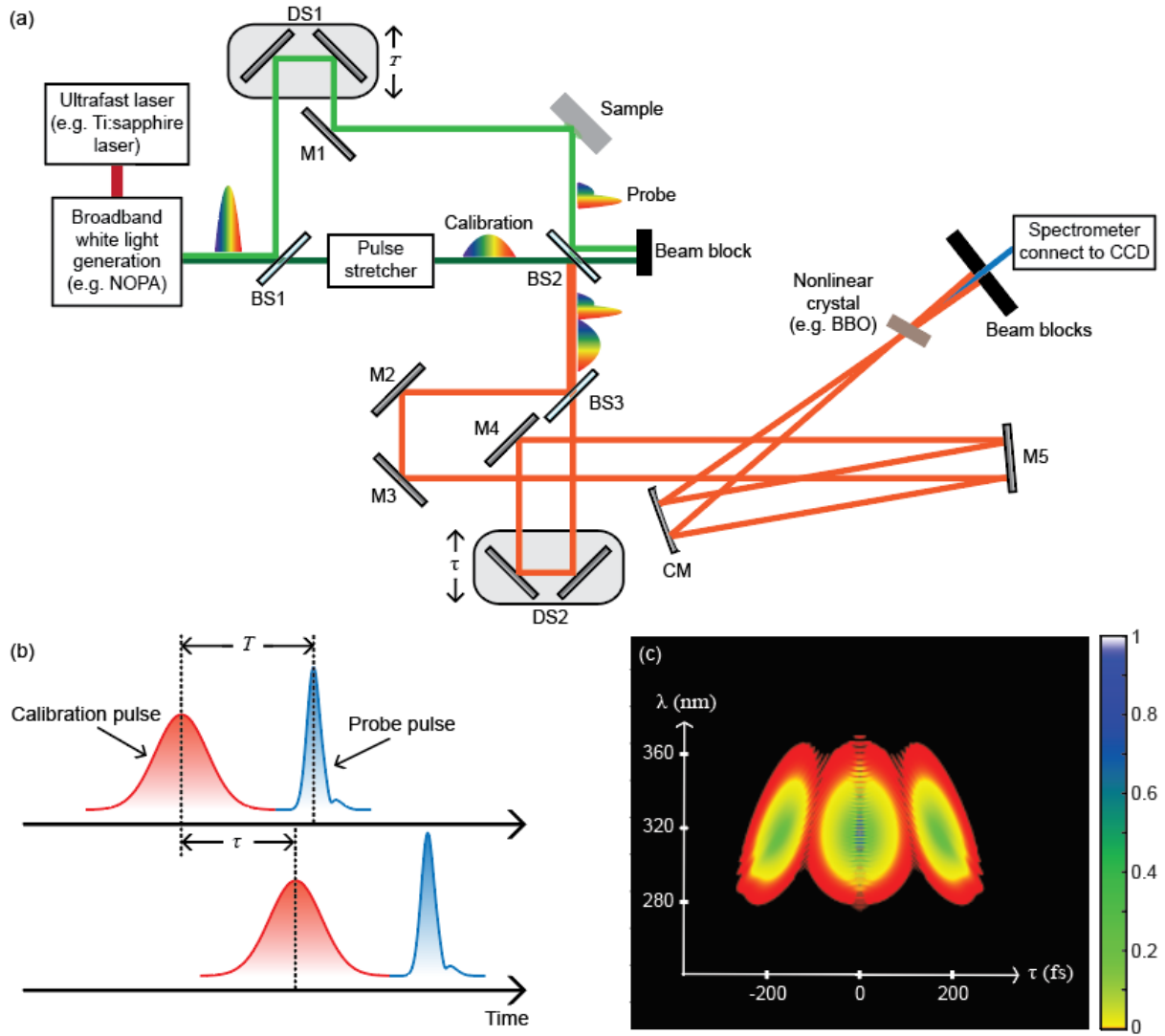

**Figure 1** (a) Schematic of interferometric-FROG spectroscopy, BS1-BS3, beamsplitters; DS1-DS2, delay stages; M1-M4, mirrors; CM, concave mirror. (b) Temporal diagram of double-pulse trains with a constant inter-pulse delay $T$ and variable inter-train delay $\tau$ for generating 2D FROG trace in (c) showing the resultant quasi-triple-peak trace with diagonal fringes activating phase-locking mechanism.

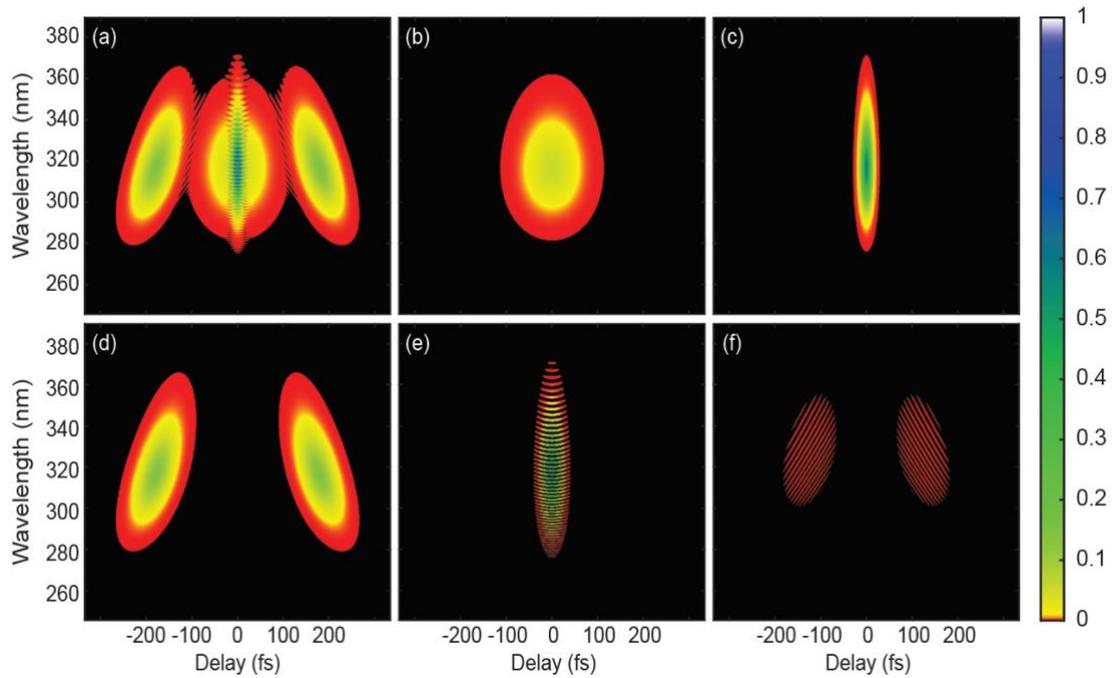

**Figure 2** (a) The simulated quasi-triple-peak FROG trace and (b-d) the decomposition analysis showing individual parts constituting FROG trace in (a) in order of calibration pulse with chirp, near-transform-limited probe pulse as well as nonlinear cross-interaction of calibration and probe pulse, by neglecting effect of spectral interference. (e-f) Decomposition analysis of contribution by spectral interference that is neglected in panel (b-d).

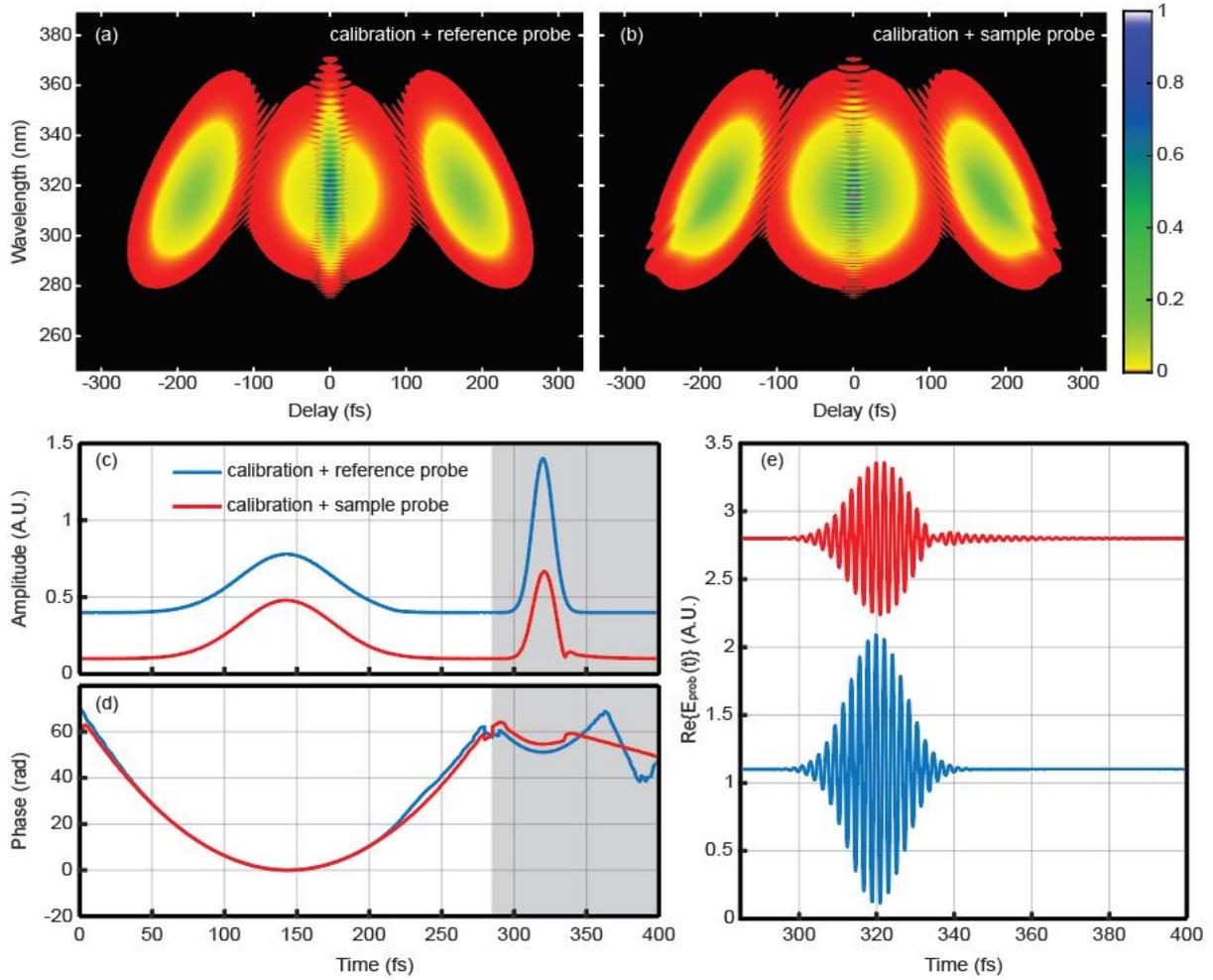

**Figure 3** Computed FROG traces for (a) calibration-reference-pulse train and (b) calibration-sample-pulse train. FROG retrieval algorithm generates time-dependent amplitude (c) and phase (d) of electric fields of the calibration-reference-pulse train (blue) and the calibration-sample-pulse train (red) of $WS_2$, respectively. Both reference and sample pulse fields (shaded regions) are extracted and then multiplied with carrier wave term for reconstruction of time-domain signals shown in (e), red for sample pulse and blue for reference pulse, as required in standard time-domain spectroscopy.

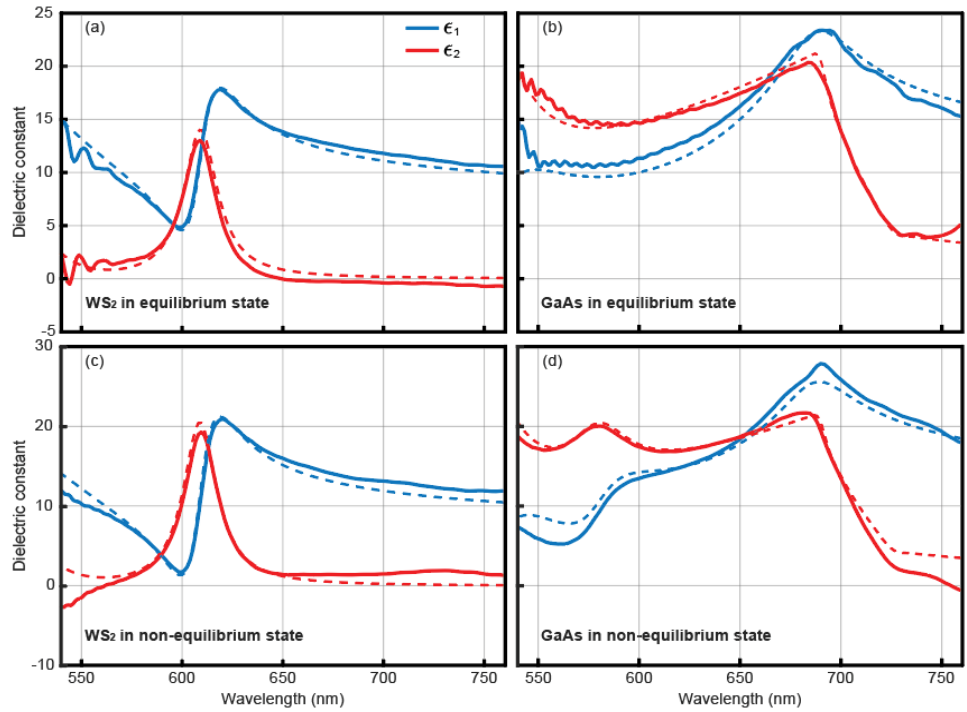

**Figure 4** Real (blue) and imaginary part (red) of dielectric functions of equilibrium and non-equilibrium state of WS$_2$ (a, c) and modulated-GaAs (b, d) are extracted in the visible range using the interferometric FROG-augmented time-domain spectroscopy (solid lines), in comparison with the expected value computed by the analytical model (dashed lines).

# Asymmetric double-pulse interferometric frequency-resolved optical gating for visible-wavelength time-domain spectroscopy: supplemental material

**S1. Explanation to Eq. (1)**

When the complex time-dependent sample electric field, $E(t)$, interacts with its delayed replica, $E(t-\tau)$, in a nonlinear crystal for the second-harmonic generation (SHG), the output time-dependent signal field can be written as:

$$E_s(t,\tau) \propto E(t)E(t-\tau) \tag{S1}$$

The frequency-dependent signal field, $\tilde{E}_s(\omega,\tau)$, can be obtained by Fourier transforming $E_s(t,\tau)$ with respect to time, $t$. This is the integration term contributing to the SHG FROG trace in Eq. (1). The 2D FROG trace is generated in frequency domain through the nonlinear process at varying delay, $\tau$,

$$S_{FROG_\omega}(\omega,\tau) \propto \left|\tilde{E}_s(\omega,\tau)\right|^2 \tag{S2}$$

As most spectrometers only perform measurement in wavelength, $\lambda$, rather than frequency, we need to related the same observable in two different domains by considering conservation of energy [1], we have:

$$\begin{aligned}
\int_{-\infty}^{+\infty} S_{FROG_\omega}(\omega,\tau)\,d\omega &= \int_{-\infty}^{+\infty} S_{FROG_\omega}\left(\frac{2\pi c}{\lambda},\tau\right) d\left(\frac{2\pi c}{\lambda}\right) \\
&= \int_{-\infty}^{+\infty} \frac{2\pi c}{\lambda^2} \times S_{FROG_\omega}\left(\frac{2\pi c}{\lambda},\tau\right) d\lambda \qquad (S3) \\
&= \int_{-\infty}^{+\infty} S_{FROG_\lambda}(\lambda,\tau)\,d\lambda
\end{aligned}$$

As a result, the additional wavelength-dependent factor, $2\pi c/\lambda^2$, is multiplied to convert the frequency-dependent FROG trace into the appropriate data format readable by standard FROG retrieval algorithm.

**S2. Mathematical interpretation of phase-locking mechanism**

Pulsed electric field can generally be described in form of a complex-valued time-dependent quantity, $E(t) = a(t)e^{i(\omega_0 t + \varphi)}$, where $a$ is the real-valued pulse envelope, $\omega_0$ is carrier wave frequency and $\varphi(t)$ is the time-dependent phase. Note that the time-dependent phase can be expanded as $\varphi(t) = \bar{\varphi}(t) + \varphi_0$, where $\bar{\varphi}(t) = \sum_{n=1}^{+\infty} \varphi_n t^n$ and $\varphi_0$ is the zero-order absolute phase. In standard SHG FROG retrieval, zero-order phase ambiguity is constantly present. To explicitly extract term $\varphi_0$ from the time-dependent signal electric field, it follows

$$E_s(t,\tau) \propto E(t)E(t-\tau)$$
$$= a(t)e^{i(\omega_0 t+\bar{\varphi}+\varphi_0)}a(t-\tau)e^{i(\omega_0 t-\omega_0\tau+\bar{\varphi}'+\varphi_0)}$$
$$= a(t)a(t-\tau)e^{i(2\omega_0 t-\omega_0\tau+\bar{\varphi}+\bar{\varphi}')}e^{2i\varphi_0} \quad (S4)$$
$$= A(t,\tau)e^{2i\varphi_0}$$

where $\bar{\varphi}'$ is $\bar{\varphi}$ with an inter-pulse-train delay ($\tau$) expandable to orders higher than zero ($n \geq 1$) and $A(t,\tau) = a(t)a(t-\tau)e^{i(2\omega_0 t-\omega_0\tau+\bar{\varphi}+\bar{\varphi}')}$. As absolute phase $\varphi_0$ is a real-valued constant, the zero-order phasor, $e^{i\varphi_0}$, remains unchanged when Fourier transforming between the time and frequency domain. The time-dependent FROG signal field, $E_s(t,\tau)$, should be Fourier transformed into frequency domain as typically shown in standard FROG trace, $\tilde{E}_s(\omega,\tau) = \tilde{A}(\omega,\tau)e^{2i\varphi_0}$. Following from Eq. (S2), 2D FROG trace results from taking absolute square of $\tilde{E}_s(\omega,\tau)$ so that absolute zero-order term ($e^{i\varphi_0}$) vanishes, i.e., phase retrieval using FROG signal of stand-alone pulse is inherently imbedded with ambiguity in $\varphi_0$. To retrieve phase information to the lowest possible order, we therefore propose asymmetric double-pulse scheme to lock the zero-order absolute phase difference of two pulses for elimination of ambiguity, a critical step toward successful phase sensitive spectroscopy at optical frequencies, e.g., visible and near-IR.

Mathematically, constituent electric field of the asymmetric double-pulse train is a superposition of calibration pulse, $E_{cal}(t)$, and probe pulse, $E_{pr}(t)$, with a fixed time delay ($T$). Its SHG FROG signal field can be written in the following way using Eq. (S1):

$$E_s'(t,\tau,T) \propto [E_{cal}(t) + E_{pr}(t-T)] \times [E_{cal}(t-\tau) + E_{pr}(t-\tau-T)]$$
$$= E_{cal}(t)E_{cal}(t-\tau) + E_{pr}(t-T)E_{pr}(t-\tau-T) + E_{cal}(t)E_{pr}(t-\tau-T)$$
$$+ E_{pr}(t-T)E_{cal}(t-\tau) \quad (S5)$$
$$= E_{cal}(t,\tau) + E_{pr}(t-T,\tau) + E_{cal-pr}(t,\tau,T) + E_{pr-cal}(t,\tau,T)$$

The first two terms, $E_{cal}(t,\tau)$ and $E_{pr}(t-T,\tau)$, respectively correspond to the stand-alone FROG signal fields of self-interacting calibration and probe pulses, whereas the third term, $E_{cal-pr}(t,\tau,T)$ (fourth term, $E_{pr-cal}(t,\tau,T)$) corresponds to cross interaction of the calibration (probe) pulse and the delayed probe (calibration) pulse inside nonlinear crystal. The asymmetric double-pulse train yields a quasi-triple-peak FROG trace that can be decomposed into four regimes shown in Fig. S1 in accordance with Eq. (1). As in standard FROG trace, the symmetric characteristic about the zero-delay line in the SHG FROG trace is well respected in the double-pulse configuration.

When the inter-pulse-train delay, $\tau$, approaches zero in *region A*, i.e., in case of near-perfect temporal overlap, the calibration and probe pulses interact with their respective replicas only, without cross interaction, forcing both $E_{cal-pr}(t,\tau,T)$ and $E_{pr-cal}(t,\tau,T)$ to vanish (Fig. S1). The FROG trace can be simplified as the absolute square of superposition of the individual FROG signal field of calibration and probe pulse shown below:

$$S_A(\omega,\tau,T) \propto \left|\int_{-\infty}^{+\infty}[E_{cal}(t)E_{cal}(t-\tau) + E_{pr}(t-T)E_{pr}(t-\tau-T)]e^{-i\omega t}\,dt\right|^2$$

$$= |\tilde{E}_{cal-cal}(\omega,\tau) + \tilde{E}_{pr-pr}(\omega,\tau,T)|^2 \qquad (S6)$$

$$= S_{cal}(\omega,\tau) + S_{pr}(\omega,\tau,T) + S_{f1}(\omega,\tau,T)$$

The first and second terms of Eq. S6 are generated by the calibration pulse, $S_{cal}(\omega,\tau) = |\tilde{E}_{cal-cal}(\omega,\tau)|^2$, (Fig. 2b) and the probe pulse, $S_{pr}(\omega,\tau,T) = |\tilde{E}_{pr-pr}(\omega,\tau,T)|^2$, (Fig. 2c) respectively, representing the stand-alone SHG FROG traces. Thus, the corresponding absolute phases of calibration ($\varphi_{0_{cal}}$) and probe ($\varphi_{0_{pr}}$) pulses vanish. The third term demonstrates spectral interference between the FROG signals of calibration and probe pulse. In time domain, stand-alone calibration and probe pulse FROG signal fields ($E_{cal-cal}(t,\tau)$ and $E_{pr-pr}(t-T,\tau)$) read

$$E_{cal-cal}(t,\tau) = E_{cal}(t)E_{cal}(t-\tau)$$

$$= a_{cal}(t)e^{i(\omega_0 t + \bar{\varphi}_{cal} + \varphi_{0_{cal}})} a_{cal}(t-\tau) e^{i(\omega_0 t - \omega_0 \tau + \bar{\varphi}'_{cal} + \varphi_{0_{cal}})}$$

$$= a_{cal}(t)a_{cal}(t-\tau)e^{i[\omega_0(2t-\tau) + \bar{\varphi}_{cal} + \bar{\varphi}'_{cal}]} e^{2i\varphi_{0_{cal}}} \qquad (S7)$$

$$= A_{cal-cal}(t,\tau)e^{2i\varphi_{0_{cal}}}$$

$$E_{pr-pr}(t-T,\tau) = E_{pr}(t-T)E_{pr}(t-\tau-T)$$

$$= a_{pr}(t-T)e^{i(\omega_0 t - \omega_0 T + \bar{\varphi}_{pr} + \varphi_{0_{pr}})} a_{pr}(t-\tau-T) e^{i(\omega_0 t - \omega_0 \tau - \omega_0 T + \bar{\varphi}'_{pr} + \varphi_{0_{pr}})}$$

$$= a_{pr}(t-T)a_{pr}(t-\tau-T)e^{i[\omega_0(2t-2T-\tau) + \bar{\varphi}_{pr} + \bar{\varphi}'_{pr}]} e^{2i\varphi_{0_{pr}}} \qquad (S8)$$

$$= A_{pr-pr}(t-T,\tau)e^{2i\varphi_{0_{pr}}}$$

where $a_{cal}$ ($a_{pr}$) is the envelope of the calibration (probe) pulse, $\bar{\varphi}_{cal}$ ($\bar{\varphi}_{pr}$) is the time-dependent phase expandable in powers of time at orders higher than zero ($n \geq 1$), $A_{cal-cal}$ ($A_{pr-pr}$) has the same magnitude as $E_{cal-cal}$ ($E_{pr-pr}$) but with a phase difference and $A_{pr-pr}(t) = a_{pr}(t)a_{pr}(t-\tau)e^{i[\omega_0(2t-\tau)+\bar{\varphi}_{pr}+\bar{\varphi}'_{pr}]}$. The third term in Eq. S6 corresponds to the horizontal fringes, jointly determined by intra-pulse-train delay $\omega T$ and calibration-probe phase difference at zero order ($\Delta\varphi_0 = \varphi_{0_{cal}} - \varphi_{0_{pr}}$). It can be written as the following:

$$S_{f1}(\omega,\tau,T) = \tilde{E}_{cal-cal}(\omega,\tau)\tilde{E}^*_{pr-pr}(\omega,\tau,T) + c.c.$$

$$= \int_{-\infty}^{+\infty} E_{cal-cal}(t,\tau)e^{-i\omega t}\,dt \left[\int_{-\infty}^{+\infty} E_{pr-pr}(t-T,\tau)e^{-i\omega t}\,dt\right]^* + c.c.$$

$$= \int_{-\infty}^{+\infty} A_{cal-cal}(t,\tau)e^{2i\varphi_{0_{cal}}}e^{-i\omega t}\,dt \qquad (S9)$$

$$\times \left[\int_{-\infty}^{+\infty} A_{pr-pr}(t-T,\tau)e^{2i\varphi_{0_{pr}}}e^{-i\omega t}\,dt\right]^* + c.c.$$

$$= \tilde{A}_{cal-cal}(\omega,\tau)e^{2i\varphi_{0_{cal}}}dt\left[\tilde{A}_{pr-pr}(\omega,\tau)e^{-i\omega T}e^{2i\varphi_{0_{pr}}}\right]^* + c.c.$$

$$= \tilde{A}_{cal-cal}(\omega,\tau)\tilde{A}^*_{pr-pr}(\omega,\tau)e^{i[\omega T+2(\varphi_{0_{cal}}-\varphi_{0_{pr}})]} + c.c.$$

$$= 2|\tilde{A}_{cal-cal}(\omega,\tau)\tilde{A}_{pr-pr}(\omega,\tau)|\cos(\omega T + 2\Delta\varphi_0 - \phi_{f1})$$

$$= 2\sqrt{S_{cal}(\omega,\tau)}\sqrt{S_{pr}(\omega,\tau)}\cos(\omega T + 2\Delta\varphi_0 - \phi_{f1})$$

where $\sqrt{S_{cal}} = |\tilde{A}_{cal-cal}|$ ($\sqrt{S_{pr}} = |\tilde{A}_{pr-pr}|$) is the magnitude of $\tilde{E}_{cal-cal}$ ($\tilde{E}_{pr-pr}$) pulse in frequency domain, $\phi_{f1}$ is the spectral phase difference, at order higher than zero, between $\tilde{A}_{cal-cal}$ and $\tilde{A}_{pr-pr}$. By only considering the horizontal fringes term ($S_{f1}$), it gives similar results as the spectral interferometry (SI), $S_{SI,fringes}(\omega,\tau) = 2\sqrt{S_{cal}(\omega)}\sqrt{S_{pr}(\omega)}\cos[\Delta\varphi(\omega) + \omega T]$, e.g., TADPOLE combined SI and FROG measurement [2–3]. Although both methods use asymmetry double-pulse train to acquire phase sensitivity, SI has limited spectral resolution and requires very high precision in measuring the time difference between the calibration and probe pulse ($T$), which could become inextricably frequency dependent due to group velocity dispersion. In contrast, our proposed method can circumvent the difficulty by shifting work of tracking phase difference, in particular, at zero order to diagonal fringes to be discussed below in region C.

In the non-trivial *region C*, $\tau$ is getting large enough for calibration pulse in one train to simultaneously interact with the tail of its own replica and full width of probe pulse in the other double-pulse train. Such interaction generates double FROG signal field in time domain (Fig. S1b). The resultant FROG trace becomes:

$$S_C(\omega,\tau,T) \propto \left|\int_{-\infty}^{+\infty} E_{cal}(t)[E_{cal}(t-\tau) + E_{pr}(t-\tau-T)]e^{-i\omega t}\,dt\right|^2$$

$$= |\tilde{E}_{cal-cal}(\omega,\tau) + \tilde{E}_{cal-pr}(\omega,\tau,T)|^2 \tag{S10}$$

$$= S_{cal}(\omega,\tau) + S_{cal-pr}(\omega,\tau,T) + S_{f2}(\omega,\tau,T)$$

The first term, $S_{cal}(\omega,\tau) = |\tilde{E}_{cal-cal}(\omega,\tau)|^2$, again represents the intensity profile of the stand-alone SHG FROG trace of the calibration pulse. The second term corresponds to calibration-probe pulse interaction, giving rise to off-center peak, $S_{cal-pr}(\omega,\tau,T) = |\tilde{E}_{cal-pr}(\omega,\tau,T)|^2$, in the quasi-triple-peak FROG trace. It is the intensity of the frequency-dependent signal field generated by temporal overlap part of full probe pulse and a fraction of calibration pulse in nonlinear crystal. The mathematical interpretation can be obtained from reviewing time-dependent signal field of the cross-interaction of calibration and probe pulse:

$$\begin{aligned}E_{cal-pr}(t,\tau,T) &= E_{cal}(t)E_{pr}(t-\tau-T)\\ &= a_{cal}(t)e^{i(\omega_0 t+\bar{\varphi}_{cal}+\varphi_{0_{cal}})}a_{pr}(t-\tau-T)e^{i(\omega_0 t-\omega_0\tau-\omega_0 T+\bar{\varphi}'_{pr}+\varphi_{0_{pr}})}\\ &= a_{cal}(t)a_{pr}(t-\tau-T)e^{i[\omega_0(2t-\tau-T)+\bar{\varphi}_{cal}+\bar{\varphi}'_{pr}]}e^{i(\varphi_{0_{cal}}+\varphi_{0_{pr}})}\\ &= A_{cal-pr}(t,\tau,T)e^{i(\varphi_{0_{cal}}+\varphi_{0_{pr}})}\end{aligned} \tag{S11}$$

where $A_{cal-pr}$ is the amplitude term carrying time-dependent amplitude and phase, to be Fourier transformed as a whole for signal field in frequency domain. Similar to the stand-alone FROG trace, two phase terms ($\varphi_{0_{cal}}$ and $\varphi_{0_{pr}}$) remain unchanged after Fourier transforming into frequency domain, $\tilde{E}_{cal-pr}(\omega,\tau,T)$, but vanish after taking absolute square for FROG trace, $S_{cal-pr}(\omega,\tau,T) = |\tilde{E}_{cal-pr}(\omega,\tau,T)|^2$. The zero-order phase difference, $\Delta\varphi_0$, will survive in term ($S_{f2}$), due to interference between $\tilde{E}_{cal-cal}(\omega,\tau)$ and $\tilde{E}_{cal-pr}(\omega,\tau,T)$, and subsequently create diagonal interference fringes. Using Eq. (S7) and Eq. (S11), the term responsible for tilted interference fringes is as follows:

$$\begin{aligned}
S_{f2}(\omega,\tau,T) &= \tilde{E}_{cal-cal}(\omega,\tau)\tilde{E}^*_{cal-pr}(\omega,\tau,T) + c.c. \\
&= \int_{-\infty}^{+\infty} E_{cal-cal}(t,\tau)e^{-i\omega t}\,dt \left[\int_{-\infty}^{+\infty} E_{cal-pr}(t,\tau,T)e^{-i\omega t}\,dt\right]^* + c.c. \\
&= \int_{-\infty}^{+\infty} A_{cal-cal}(t,\tau)e^{2i\varphi_{0_{cal}}}e^{-i\omega t}\,dt \\
&\quad \times \left[\int_{-\infty}^{+\infty} A_{cal-pr}(t,\tau,T)e^{i(\varphi_{0_{cal}}+\varphi_{0_{pr}})}e^{-i\omega t}\,dt\right]^* + c.c. \\
&= \tilde{A}_{cal-cal}(\omega,\tau)e^{2i\varphi_{0_{cal}}}dt\left[\tilde{A}_{cal-pr}(\omega,\tau,T)e^{i(\varphi_{0_{cal}}+\varphi_{0_{pr}})}\right]^* + c.c. \\
&= \tilde{A}_{cal-cal}(\omega,\tau)\tilde{A}^*_{cal-pr}(\omega,\tau,T)e^{i(\varphi_{0_{cal}}-\varphi_{0_{pr}})} + c.c. \\
&= 2|\tilde{A}_{cal-cal}(\omega,\tau)\tilde{A}_{cal-pr}(\omega,\tau,T)|\cos(\Delta\varphi_0 - \phi_{f2}) \\
&= 2\sqrt{S_{cal}(\omega,\tau)}\sqrt{S_{cal-pr}(\omega,\tau,T)}\cos(\Delta\varphi_0 - \phi_{f2})
\end{aligned} \quad (S12)$$

where $\sqrt{S_{cal-pr}} = |\tilde{A}_{cal-pr}|$ is the magnitude of $\tilde{E}_{cal-pr}$, and $\phi_{f2}$ is the FROG-retrievable first- (and above) order spectral phase difference between $\tilde{A}_{cal-cal}$ and $\tilde{A}_{cal-pr}$. In stark contrast to term $S_{f1}$, the expression above ($S_{f2}$) says $T$ is *decoupled* from sinusoidal modulation term and get absorbed into $\tilde{E}_{cal-pr}$, which can be precisely retrieved $E'(t,\tau,T) = E_{cal}(t) + E_{pr}(t,T)$, using standard FROG algorithm, even if $T$ is frequency dependent after propagating short pulse through dispersive material. To this point, a two-dimensional search of $T$ and $\Delta\varphi_0$ with uncertainty as in spectral interference (SI) experiment has been reduced to one-dimensional search of the interlocked phase difference $\Delta\varphi_0$ that now can be precisely determined, i.e., the phase-locking mechanism preserves the phase difference of the calibration and probe pulse, $\Delta\varphi_0$ and encodes it into the FROG trace.

In this new scheme, the time-dependent electric field of the asymmetric double pulse, $E'(t,\tau,T)$, can be directly retrieved using standard FROG algorithm. The individual absolute phases ($\varphi_{0_{cal}}$ and $\varphi_{0_{pr}}$) still remain ambiguous, but their difference $\Delta\varphi_0$ is now completely ambiguity-free, sufficient for successful reconstructing time-domain signal, as detailed in section S4. This is called the phase-locking mechanism enabled by fringe-generating term. Most importantly, we removed uncertainty in the frequency-dependent inter-pulse delay ($\omega T$), detrimental to any meaningful spectroscopy measurement of practical use.

## S3. Interpretation of other regions in quasi-triple-peak FROG trace

In section S2, origins of horizontal fringes in region A (Eq. S9) and the diagonal fringes in region C (Eq. S12) were accounted for by spectral interference effect, dictated by spectral phase factor respectively in the sinusoidal term. Now we move on to discussing region B and D of the quasi-triple-peak FROG trace.

In region B, the probe pulse has limited contribution to the FROG trace due to its significantly shorter pulse duration than inter-pulse delay $T$, and only $E_{s,cal}(t,\tau) = E_{cal}(t)E_{cal}(t-\tau)$ in Eq. (S5) survives. The FROG trace of region B can be written in the following way:

$$S_B(\omega,\tau,T) \propto \left| \int_{-\infty}^{+\infty} E_{cal}(t)E_{cal}(t-\tau)e^{-i\omega t}\,dt \right|^2 \qquad (S13)$$
$$= S_{cal}(\omega,\tau)$$

It represents a standard SHG FROG trace of the self-interacting calibration pulse with ambiguity in $\varphi_{0_{cal}}$.

In region D, two pulse trains are separated apart from each other, so that interaction of delayed probe pulse in one train to calibration pulse in the other train predominantly contributes to the FROG trace, in the absence of spectral interference effect. The corresponding FROG signal reads:

$$S_D(\omega,\tau,T) \propto \left| \int_{-\infty}^{+\infty} E_{cal}(t)E_{pr}(t-\tau)e^{-i\omega t}\,dt \right|^2$$
$$= \left| \tilde{E}_{cal-pr}(\omega,\tau,T) \right|^2 \qquad (S14)$$
$$= S_{cal-pr}(\omega,\tau,T)$$

## S4. Spectroscopic analysis using retrieved time domain fields

To examine the capability of asymmetric double-pulse interferometric FROG (ADI-FROG) in the visible regime, numerical simulation was conducted using two semiconducting materials, WS$_2$ and modulated-GaAs (bandgap redshifted by 285 nm), in both static and non-equilibrium states. A linearly chirped probe pulse with a time-dependent electric field of $E_{ref}(t) = e^{-(2\ln 2)t^2/[T_{TL}^2+i(\ln 2)\beta]}e^{-i2\pi ct/\lambda_0}$ is used, where $T_{TL}$ is the transform-limited pulse duration of 7.8 fs, $\beta$ is a linear chirp parameter of 100 fs$^2$, $c$ is the speed of light and $\lambda_0$ is the central wavelength of 635 nm. The corresponding probe pulse covers spectral range from 550 nm to 750 nm. As required in standard time-domain spectroscopy, interacting pristine Gaussian probe pulse with material of unity reflection, e.g., silver, and with sample surface would give what we respectively call *reference* and *sample pulse.*

The sample pulse, in principle, could be experimentally obtained by reflecting a pristine probe pulse off the sample surface (WS$_2$ or modulated-GaAs) at normal incidence. At numerical simulation level as in this work, we Fourier-transform time-domain reference pulse, $E_{ref}(t)$, into frequency domain and then plug into Fresnel equation to compute sample electric field, using analytical complex dielectric function of the sample, $\tilde{\epsilon}(\omega) = \epsilon_1 + i\epsilon_2$, where $\epsilon_1$ and $\epsilon_2$ are real and imaginary parts of the complexed-valued dielectric function, respectively.

$$\tilde{E}_{sam}(\omega) = \tilde{E}_{ref}(\omega) \frac{1 - \sqrt{\tilde{\epsilon}(\omega)}}{1 + \sqrt{\tilde{\epsilon}(\omega)}} \tag{S15}$$

The analytical complex dielectric function $\epsilon(\omega)$ describes spectroscopic characteristics of the sample, and generates time-domain signal of sample pulse $E_{sam}(t)$ by inverse Fourier transform. To emulate proposed experiment, $E_{sam}(t)$ would be combined with calibration pulse ($E_{cal}(t)$) to synthesize asymmetric double-pulse train as the input of FROG measurement. Subsequently, feasibility of FROG-based method described herein will have to be verified by the level of match between *analytical dielectric function* and that yielded from the protocol: 1. Generating phase-locked FROG traces, one for calibration-sample pulse pair and the other for calibration-reference pulse pair; 2. Retrieval and reconstruction of sample & reference pulse in time domain; 3. Fourier transforming and plugging reconstructed signals of sample and reference pulses back into Fresnel equation for numerical simulation output $\tilde{\epsilon}_{num}(\omega)$ to be compared with its analytical counterpart.

To simulate the non-equilibrium dynamics, additional Lorentz oscillator is purposely imposed to static state dielectric function of the sample $\tilde{\epsilon}(\omega)$ for transient state dielectric function below

$$\tilde{\epsilon}'(\omega) = \tilde{\epsilon}(\omega) + \frac{\omega_p^2}{(2\pi c/\lambda_r)^2 - \omega^2 - i\gamma\omega} \tag{S16}$$

where $\omega_p$ is the plasma frequency set at 1423.3 THz, $\lambda_r$ is the resonant wavelength of 640 nm, and $\gamma$ is the damping rate of 100 THz. As did in static state analysis, the frequency-dependent sample pulse field in non-equilibrium state can be calculated by replacing $\tilde{\epsilon}(\omega)$ with $\tilde{\epsilon}'(\omega)$ and then using Eq. (S15) and Eq. (S16). Again, inverse Fourier transform is conducted to obtain the time-domain sample electric field to superpose with a synchronized calibration pulse of $E_{cal}(t) = e^{-(2\ln 2)t^2/[T_{TL}^2 + i(6\ln 2)\beta]}e^{-i2\pi ct/\lambda_0}$, which possesses longer pulse duration due to a six times larger linear chirp parameter $\beta$ than that in the probe pulse. The time delay between two pulses, $T$, is set at 180 fs, which is about 2.7 times longer than the sum of calibration and probe pulse durations. The asymmetric double-pulse train is superposition of a leading calibration pulse and a trailing probe pulse, to interact with either a gold reference or sample of interest. The 2D quasi-triple-peak FROG traces of a double-pulse train, comprised of calibration pulse and reference/sample (WS$_2$ and modulated GaAs) pulse probing either equilibrium or non-equilibrium state of the sample, can be generated using Eq. (1) through Eq. (3) (Fig. S2 a-e). To reconstruct the time-domain signal, we use a phase-retrieval software (Fig. S2 f-j), Femtosoft FROG 3.2.4 (Femtosoft) developed by Dr. K. W. DeLong in the early 1990s at Sandia National Laboratories, mainly to implement iterative phase retrieval algorithm based on generalized projections [4–5]. The uncertainty, evaluated by subtracting the original computed FROG trace from the reconstructed one, is found less than 1.2 % of peak intensity of the original traces, adequate for accurate retrieval of the asymmetric double-pulse train.

To precisely obtain complex dielectric function, calibration pulses in two sets of retrieval, i.e. calibration-reference pulse train and calibration-sample pulse train, are normalized in both amplitude and phase to one another. This is motivated by the fact that calibration pulses from two independent double-pulse trains are identical. In case of asymmetric double pulse train, both time-domain amplitude and phase of the double-pulse train can be completely and directly retrieved. For a specific double-pulse train, the relative phase difference between calibration and reference/sample pulse given by FROG retrieval algorithm always stages constant due to phase-locking mechanism. However, from retrieval to retrieval, random uncertainty in collective phase offset and calibration pulse amplitude still prevails. To eliminate such uncertainty, we simply force the calibration pulses in all pulse trains to be identical by amplitude normalization and phase offsetting. As a result, the calibration pulse in calibration-sample train now has the same amplitude as the one in the calibration-reference pulse train (Fig. 3c). Finally, relative difference in amplitude and phase of the trailing reference and sample pulse can be fully attributed to light-sample interaction and therefore be taken as input for a reconstruction of time-domain signals of sample and reference scans, as in time-domain spectroscopy at THz and multi-THz frequencies.

Note that the phase-locking mechanism is not able to measure the real-valued zero-order absolute phase carried by either calibration or probe pulse, but able to lock the calibration-sample absolute phase difference, $\Delta\varphi_0 = \varphi_{0_{cal}} - \varphi_{0_{pr}}$, to a fixed value within one pulse train. Although the reconstructed zero-order phases of calibration pulses ($\varphi_{0_{cal}}$) from different retrieval randomly varies due to ambiguity introduced by FROG retrieval algorithm, a random phase difference at zero order means a constant that can be readily removed by assigning a phase offset for a perfect match in time-dependent phase curves of presumably identical calibration pulses (Fig. S3). Consequently, the remaining phase difference in complex phases of reference and sample pulses will be of spectroscopic significance and free from adverse effect from trivial ambiguity in standard FROG retrieval. To this point, we say reference and sample pulse are precisely calibrated in amplitude and properly aligned in phase to each other using the above protocol.

As the final step, time-domain signal of calibration pulse in the reconstructed double pulse train is manually cut out, and we only keep time-domain signal of the probe pulse for further spectroscopic analysis. The complete electric fields of the reference, $E_{ref}(t) = a_{ref}(t)e^{i(-\omega_0 t + \varphi_{ref})}$, and sample, $E_{sam}(t) = a_{sam}(t)e^{i(-\omega_0 t + \varphi_{sam})}$, probe pulses can be obtained by incorporating the carrier wave with a central frequency of $\omega_0$, a known parameter, into the retrieved intensity envelope $a_{ref}$ ($a_{sam}$) of reference (sample) pulse as well as the reconstructed phase of reference (sample) pulse $\varphi_{ref}$ ($\varphi_{sam}$). Example time-domain electric fields are shown in Fig. 3e. Finally, the frequency-dependent fields can be obtained by taking Fourier transform of the time-domain signal. Both real and imaginary parts of the complex-valued dielectric function $\tilde{\varepsilon}(\omega)$ can be numerically calculated using Fresnel equation Eq. (S15).

**Supplementary Figures**

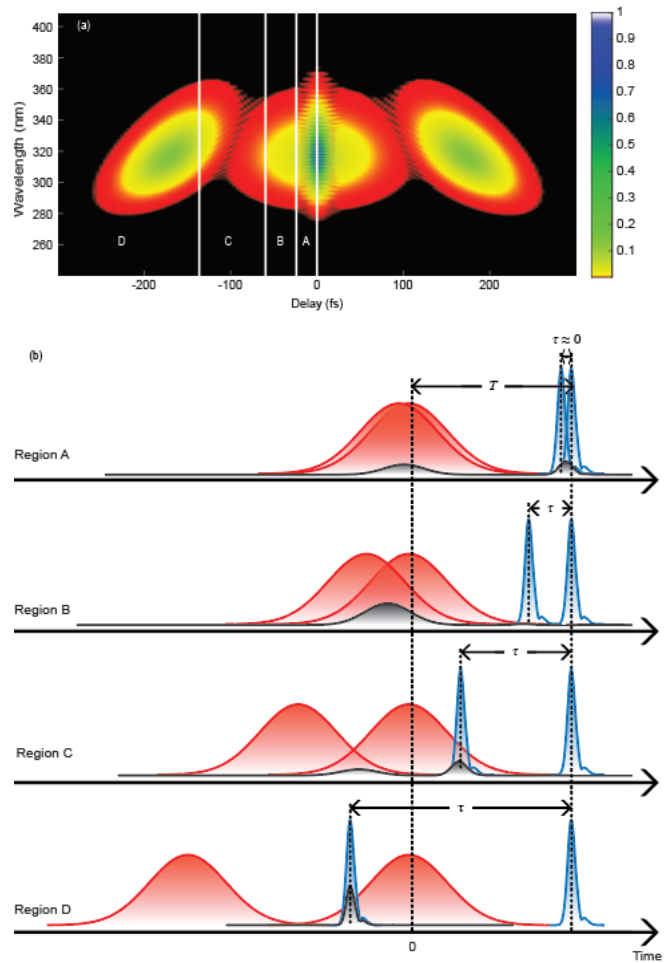

**Fig. S1** (a) Four major regions of the asymmetric double-pulse FROG trace are symmetric about the zero-delay line. Different regions of A, B, C and D are defined in order of increasing inter-pulse-train delay $\tau$. (b) Time-domain envelope of double-pulse trains (calibration, red; probe, blue) at various delays for illustrating effect of inter-pulse-delay on the resultant SHG FROG signal (black).

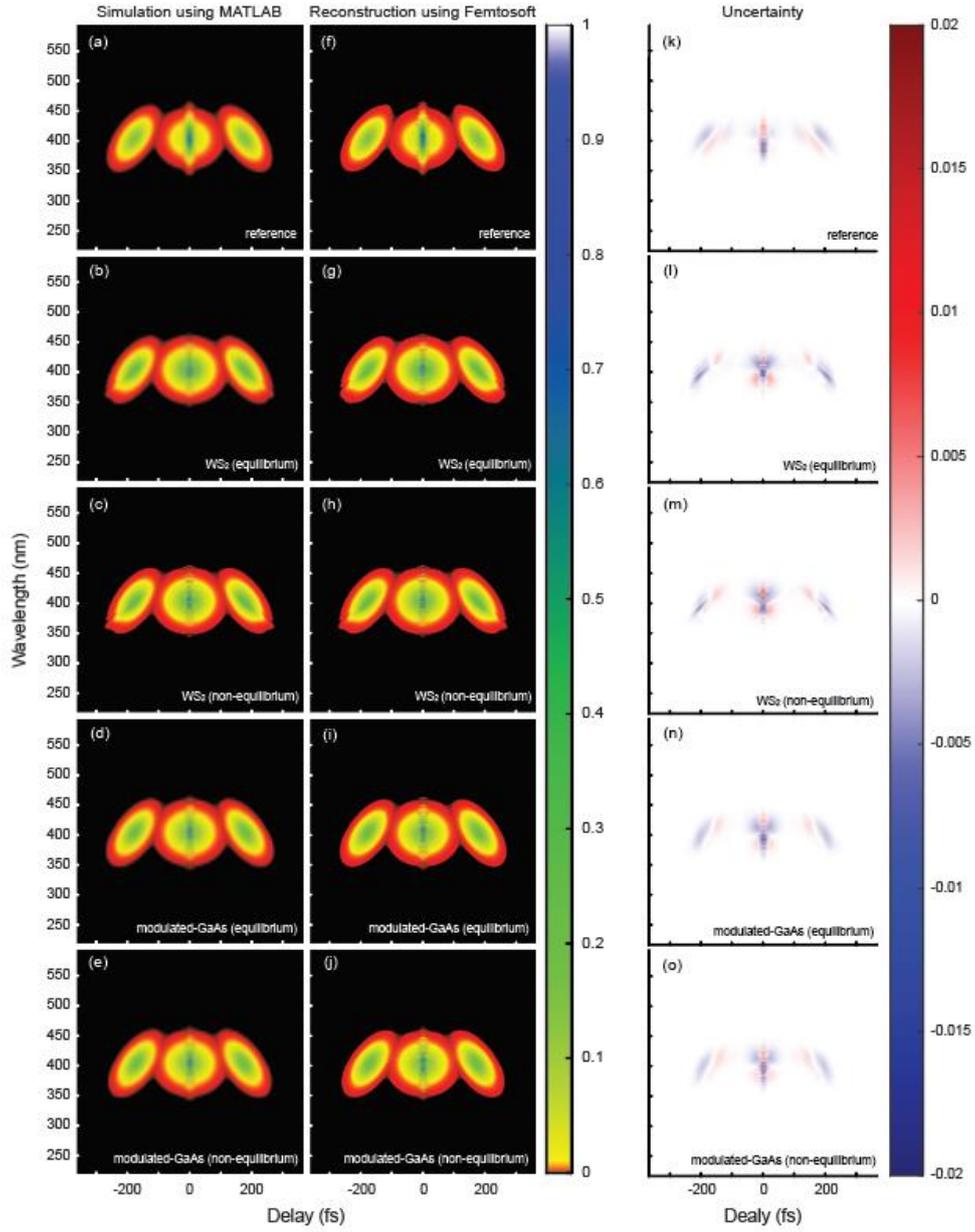

**Fig. S2.** (a)-(e) Analytically computed ADI-FROG traces of a calibration pulse superposing with a reference or sample (WS$_2$ and modulated GaAs) pulse in either equilibrium or non-equilibrium state. They mimic measured FROG traces as in real experiments. (f)-(j) Retrieved FROG traces using Femtosoft. (k)-(o) Error distribution of individual retrieval in time-wavelength space.

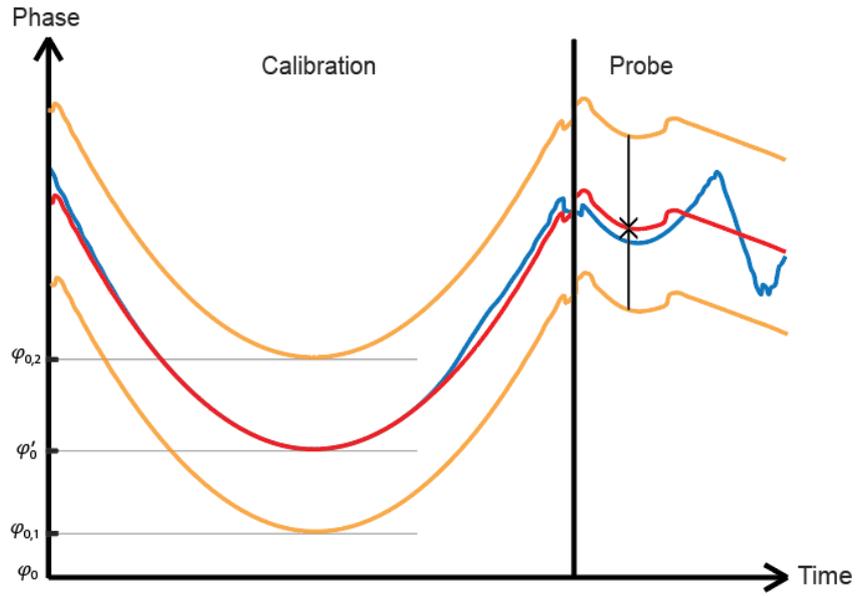

**Fig. S3** Retrieved phase of calibration-sample pulse train (red) and calibration-reference pulse train (orange) show constant difference in absolution phase $\varphi_0'$ and $\varphi_{0,1}$ (or $\varphi_{0,2}$). Note $\varphi_{0,1}$ and $\varphi_{0,2}$ represent phase curves of two independent retrievals of the same FROG traces, showing a random zero-order phase offset with respect to retrieved phase of calibration-sample pulse. This offset can be manually corrected by vertically shifting orange curves to match red. The calibrated phase curve is shown in blue color.